\documentclass[
reprint,
superscriptaddress,
showpacs,preprintnumbers,
amsmath,amssymb,
aps,prl, 
floatfix,
]{revtex4-2}

\usepackage{graphicx}
\usepackage{dcolumn}
\usepackage{bm}
\usepackage{hyperref}
\usepackage{epstopdf}
\usepackage{amsmath}
\usepackage{times}
\usepackage{natbib}

\begin{document}


\title{Light--wave Control of Non--equilibrium   Correlated States using Quantum Femtosecond Magnetism and Time--Periodic Modulation 
of Coherent Transport} 

\author{P. C. Lingos} 
\affiliation{ Department of Physics, University of Crete, 
	Box 2208,  Heraklion, Crete, 
	71003, Greece}
	
\author{M. D. Kapetanakis} 
\affiliation{Department of Physics, University of Alabama at Birmingham, Birmingham, Alabama 35294, USA}

\author{J. Wang}
\affiliation{Ames Laboratory and Department of Physics and Astronomy, Iowa State University, Ames, Iowa 50011, USA}
	
\author{I. E. Perakis}
\affiliation{Department of Physics, University of Alabama at Birmingham, Birmingham, Alabama 35294, USA}

\date{\today}

\begin{abstract}
Lightwave quantum electronics utilizes the oscillating
carrier wave of intense laser fields  to control  quantum materials properties.
 Using quantum kinetic equations of motion,
 we describe lightwave--driven nonlinear quantum transport  of  electronic spin and charge 
with simultaneous 
 quantum fluctuations
of non--collinear local spins.  During cycles of 
field oscillations,   
spin--charge  inter--atomic quantum excitations  trigger 
non--adiabatic time evolution 
of an 
antiferromagnetic  insulator state
into a 
metallic non--equilibrium state with transient magnetization.   Lightwave modulation of  
  electronic hopping  
 changes the energy landscape and 
establishes a non--thermal pathway to   laser--induced transitions  in 
 correlated systems with  strong local magnetic exchange 
  interactions. 
\end{abstract}
\maketitle

\textit{Introduction}.--
Emergent phenomena  in quantum materials  arise from competing and cooperative interactions between electronic, spin, and lattice degrees of freedom 
 \cite{review,tokura}. 
 Quasi--equilibrium (adiabatic) tuning of   multi--component order parameters and microscopic interactions,   e.g., 
by high pressure,  magnetic or electric 
fields, 
has been used  to control the  complex phase diagram. However, 
static perturbations  
  simultaneously affects other  material properties that can act against the desired effects.
 Ultrashort laser pulses provide a different route for 
manipulating  structural and electrical properties of 
quantum materials far from equilibrium. Ultrafast excitation nonlinear processes can give access 
to  metastable and prethermalized non-equilibrium states in ways not possible through quasi--equilibrium processes 
\cite{PIPT, mihail,mitrano,
aver, fausti, morris,  natmat,  
prb-sc, 
porer, 
 femtomag, prl-yang,prx, nat photon,prl_2020,cav,gedik,  
vo2,demsar1,demsar2,rini,schmitt,lingos-17}. 
Unlike for  photoexcitation  at optical frequencies, 
the advent of  intense terahertz, midinfrared, and attosecond  laser pulses  
 with few cycles of oscillation and  well--characterized
electric-field temporal  profiles  
has   opened new oppportunities   for  non--adiabatic 
 quantum tuning   
\cite{nat photon,prl_2020,prx,huber-nat,Huber-subcycle,langer,valley,prl-20,lingos-2015,shah-rev,josa,shah-prl}. 
For example, 
the lightwave electric field  can act  as an
oscillating  force to accelerate electrons in controllable trajectories
\cite{langer,valley,nat photon,prl_2020,huber-nat,Huber-subcycle,prl-20} 
Such  electronic quantum transport during cycles of lightwave or lattice coherence 
oscillations can lead to the establishment of 
 non--equilibrium or pre-thermalized  states 
prior to relaxation
\cite{natmat,nat photon,prl_2020,prl-yang,luo,floquet}.

\begin{figure*}[t]
\begin{center}
\includegraphics [scale=0.45] {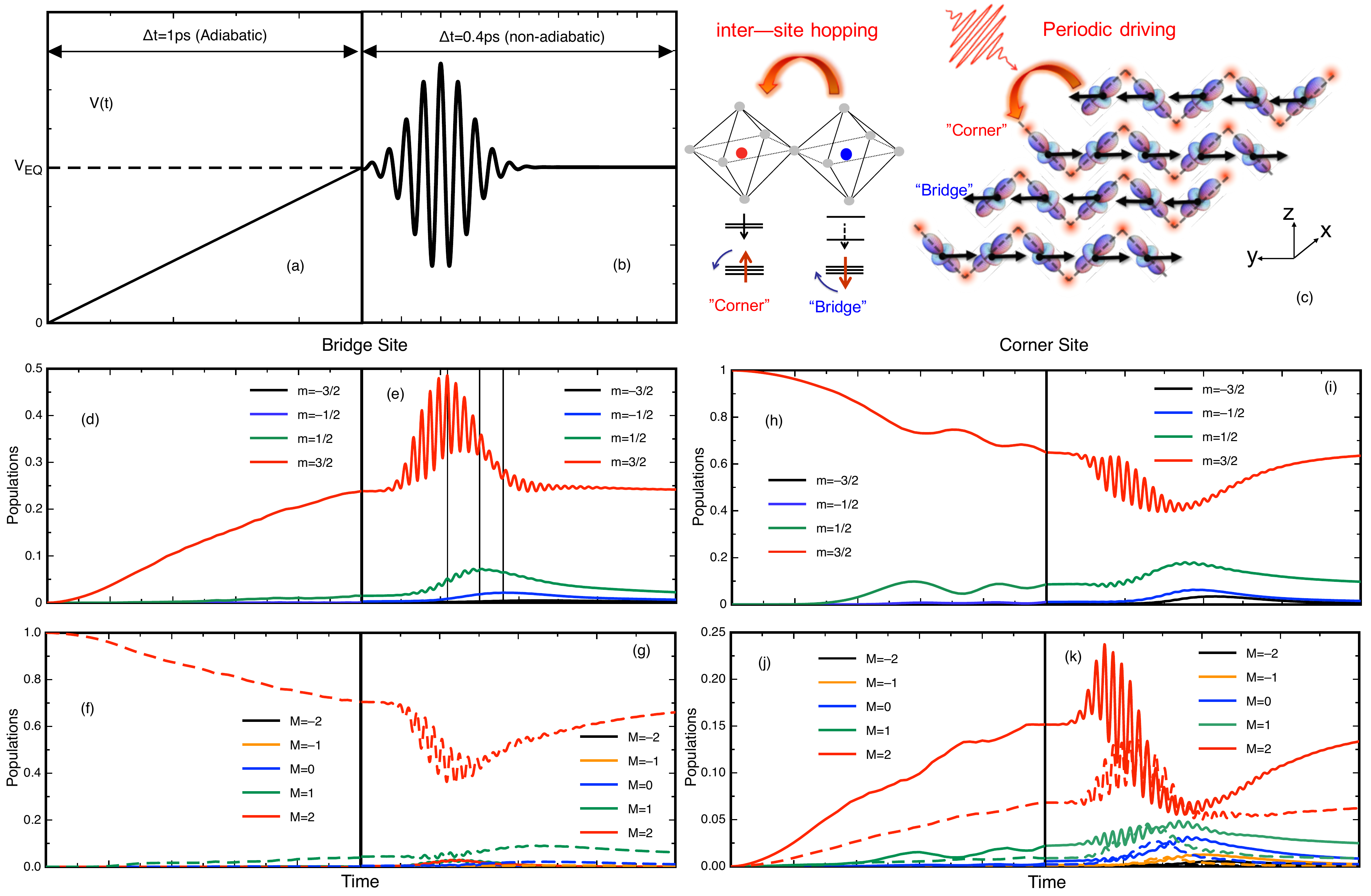}  
\caption{(Color online) The time evolution of a  CE/AFM initial state with $V$=0, (c), is driven by modulation of the inter--site hopping amplitude $V(t)$
with both adiabatic, (a),  and non--adiabatic, (b), components and 
leads to 
nonthermal  spin--dependent    populations of the local configurations at ``bridge" ($Q_B \ne 0$,
(d)--(g)) and ``corner" ($Q_C$=0, (h)--(k))  lattice sites.  
(a,b): Time dependence of $V(t)$ with  adiabatic, (a), and non--adiabatic, (b), components. The $V(t)$ profile in (b) comes from nearest neighbor hopping amplitude modulation by a multicycle 
electric field pulse with Rabi energy 
 $d_R$=$e E a$=100meV (see Supplementary).   
(c):  Schematic of the CE/AFM  initial condition. Zig--zag FM chains  consist of interchanging 
corner (red circle) and bridge ( blue circle) lattice sites. 
Neighboring  chains are AFM--coupled and are stacked in 
AFM--coupled planes along the z--axis.   
Red arrows indicate the electron hopping between AFM sites that triggers the quantum spin dynamics of main interest here, which is prohibited for classical spins when $J_H \rightarrow \infty$.  
Dynamics of the populations of the different spin local configurations at the bridge (d,e,f,g) and corner (h,i,j,k) sites 
within the adiabatic (d,f,h,j) and non--adiabatic (e,g,i,k) temporal regimes. 
The solid curves in (d--k) corresponds to the higher energy orbitals  while the lower energy ones are shown in dashed curves. 
Vertical lines in (e) indicate  a time delay in the development  of 
 different spin state populations. 
\label{fig1}}
\end{center}
\end{figure*}

In this letter,  we investigate the hypothesis 
 that lightwave--driven  quantum transport  (coherent hopping) of electrons 
between atomic sites  with non--collinear spins 
 can be used to coherently 
control   non--equilibrium transitions and transient magnetization during cycles of oscillations \cite{femtomag,prl-20,Huber-subcycle,atto}. 
 Our main focus is on the role of 
  quantum spin non--thermal  fluctuations  
driven by 
 ultrafast 
modulation of  inter--atomic
coherent electronic  hopping.
We consider the   
 strongly responsive 
spin background in a correlated magnetic system, where the 
 (Born-Oppenheimer) adiabatic approximation of classical spins 
 \cite{Dagotto,loktev,cepas}
 breaks down during ultrashort timescales. 
 By introducing  quasi--particle Hubbard operators 
and applying  a generalized--tight--binding mean field approximation \cite{Hubbard}, 
 we treat  quantum spin fluctuations 
during 
coherent electronic hopping between lattice sites 
 in the strong coupling limit of infinite on--site magnetic interaction.
Using  quantum kinetic   equations of motion
for  the Hubbard quasi--particle   density matrix
defined on a lattice,
we describe  the non-adiabatic time evolution 
of coupled spin--charge states  driven  by  lightwaves.

The ability to experimentally control coherent electron transport on  subcycle timescales  sets the stage for attosecond magnetism \cite{atto}, quantum femtosecond 
magnetism \cite{femtomag,bigot1,bigot2,2009_PRL_KAPET,chovan-prl,chovan-prb,kondo}, 
and lightwave quantum  electronics \cite{nat photon,prl_2020,huber-nat,
Huber-subcycle,langer,valley}.
The  non--thermal spin--charge--lattice
pathway studied here can initiate 
a phase transition stabilized later by the lattice \cite{morris,porer,cav,vo2,lingos-17}. Here we show that  
 non--thermal charge and spin populations and inter--site coherences can 
 change drastically the shape of the total energy landscape as compared to equilibrium
 during multi--cycle electric field oscillations. Such ``sudden" changes in the lattice forces
 prior to relaxation
 initiate  coherent  lattice displacements
 that can lead to  a phase transition and establish new non--equilibrium  states.
 We show that non--adiabatic quantum spin--charge dynamics  
evolves the AFM ground state into a 
non--equilibrium metallic 
state with transient magnetization, not possible in equilibrium.  

Our results suggest  a microscopic mechanism for  
quantum femtosecond/attosecond magnetism \cite{femtomag,atto,spin transfer,bigot2}.
In weakly correlated magnetic systems, 
it has been debated whether 
 femtosecond magnetization dynamics
arises from  adiabatic 
processes associated with electron,
spin and phonon populations, or from  
coherent processes associated with
angular momenta interacting with photoexcited electrons \cite{bigot1,bigot2}.
Here we
study  the limit of infinite  on--site magnetic exchange 
interaction and 
demonstrate  a lightwave 
driven transition 
from an AFM 
insulating ground state 
to a metallic non--equilibrium state with FM 
 correlation. The predicted 
  emergence of  nonlinear magnetization during cycles of light--wave oscillations  is achieved by simultaneous
control of electronic, magnetic, and lattice  properties, which is essential for
 lightwave quantum magneto--electronics 
 at   the ultimate speed limit.

 \textit{Quantum Kinetic Theory}.-- 
 To  model the   strong coupling of  
 spin and charge  excitations,
we consider 
composite fermion quasi--particles  created  by 
Hubbard operators \cite{Hubbard}. These Hubbard operators 
describe transitions between the multi--electron and multi--atom  local  configurations in the lattice unit cell,   
such as, e.g, the  Zhang--Rice singlet
Cu + O configuration 
in the cuprates or corresponding 
Mn + O configurations in manganese oxides \cite{loktev}.
We adopt  
 a generalized tight-binding mean field approximation  \cite{Hubbard}
 and project out the high energy upper Hubbard band
 by assuming  strong  on--site  interactions, e.g.,  Hund's rule 
\cite{Dagotto,kapet-prb,kapet-mag}. 
We consider 
a three--dimensional 
lattice with periodic boundary conditions and obtain convergence for $4\times 4\times 4=64$ lattice sites. On each site $i$, we consider  
 local spin  ${\bf S}_i$ 
 configurations $| i m \rangle$, where $S_z$=$m$=$-S \cdots S$ and 
$S$=3/2. 
Hopping of an additional itinerant electron with spin 
${\bf s}_i$
populates local   configurations $|i \alpha M \rangle$ that 
are eigenstates of the 
 total 
spin ${\bf J}_i$=${\bf S}_i$+${\bf s}_i$.
The low energy 
configurations have
$J$=$S$+1/2 and  $J_z$=$M$=$-J \cdots J$.
We also consider 
two  orbital configurations $\alpha$ and $\beta$ per lattice site, which are split in energy 
by the local lattice displacement $Q_i$, due to, e.g., 
the Jahn--Teller (JT) effect (Supplementary) \cite{Dagotto,cepas}. 
The spin local z--axis is defined by the local canting angle $\theta_i$ that defines the equilibrium spin direction at site $i$.  
$M$=$S$+1/2 and $m$=$S$ then 
correspond to  spins pointing along  $\theta_i$, 
as in the case of a classical spin  background (Supplementary  Sections S1--S2). 
For the  antiferromagnetic (AFM) ground state here, 
$\theta_i$=$0, \pi$.

The Hamiltonian is separated into on--site and 
inter--site terms, $H(t)=H_{local} + H_{hop}(t)$. 
The local states discussed above are assumed to diagonalize 
the local Hamiltonian 
$H_{local}$  in the absence of electronic  hopping. 
Important for describing the quantum spin fluctuations of interest is 
that $|i\alpha M \rangle$ 
diagonalize the FM  onsite  interaction $J_H {\bf S}_i \cdot s_i$.  
In the strong coupling limit of large 
 $J_H$, we neglect population of the high energy upper Hubbard band, $J=S-$1/2, which restricts the electronic motion
 as compared to the weak coupling limit.  
$H_{local}$ also includes the site--dependent local energy 
that depends on the  lattice distortions 
$Q_i$ (e.g., Jahn--Teller (JT) effect) and Zeeman energies 
$E_{Bi}$ due to coupling of  
a weak external field $B_{ext}$, which  breaks the symmetry and defines the  global z--direction
(Supplementary).
We thus obtain 
an energy barrier  between ``bridge" ($Q_i=Q_{B} \ne 0$) 
and ``corner" ($Q_i \approx 0$) 
 sites
leading to an insulator energy gap.

$H_{hop}$(t) describes the coherent hopping of an itinerant spin--1/2 electron 
between lattice sites. The  nearest neighbor hopping amplitude $V_{\alpha\beta}(i-j)$ is 
modulated from equilibrium by the oscillatory lightwaves
(Supplementary Sections S3--S4). 
Coherent electronic hopping from site $j$ to site $i$ 
occurs via transitions $|j \alpha M \rangle \rightarrow |j  M-1/2  \rangle$   and 
$| i m \rangle \rightarrow |i \alpha  m + 1/2 \rangle$, 
where spin 
is conserved.
We  derive and solve the real space density matrix equations of motion
defined by Hubbard operators within a mean field approximation (Supplementary).
The diagonal  density matrix 
elements,  $\rho_i^\alpha(M,t) = \langle | i \alpha M  \rangle \langle 
i \alpha M| \rangle$
and $\rho_i(m,t)= \langle | i m \rangle \langle i m | \rangle$, 
describe the spin--resolved populations of the local configurations 
at each site $i$. Coherent electronic hopping
is described by   non--diagonal density matrix elements (quantum coherences) between 
all possible pairs of lattice site $(i,j)$ configurations,  with both light--driven and quasi--equilibrium contributions. 
Below we compare  between adiabatic 
and non--adiabatic 
time evolution,
driven by slow or fast time--dependent changes in the inter--site hopping amplitude $V(t)$.

\textit{Adiabatic Dynamics}.-- 
We first consider the adiabatic time evolution of a  
CE/AFM state   with $V$=0,  
 Fig. \ref{fig1}(c),
driven by slowly varying hopping amplitude 
 $V(t)=t_{\alpha\beta}(i-j) \frac{t}{T}$.  
 $T$ is sufficiently long so that the system reaches  
a stationary state,  $\partial_t \rho$=0,  for $t > T$.
The tight binding parameters $t_{\alpha \beta}$ used here
were taken 
from Ref. 
\cite{Dagotto}, 
but may be obtained  by fitting to 
{\em ab--initio} calculations for specific materials. 
The initial   CE--AFM
charge/orbital ordered (CO/COO) state, Fig. \ref{fig1}(c),  
 consists of AFM--coupled zig--zag chains of FM-ordered 
spins 
with alternating full (bridge, total itinerant 
electron population $n$=1, energy -$E_{JT}$) and empty (corner, $n$=0, energy 0) 
sites  (CO). These  chains  are located in  identical x--y 
planes, which are AFM coupled  along the z--direction. 
In the stationary state obtained after $t \ge  T$, the populations of the 
 bridge $M=$2 and corner $m=3/2$ local configurations  have decreased, Figs \ref{fig1}(f) and (h) respectively,
 with a simultaneous increase of  configurations with $m=3/2$ and $M=$2 respectively. 
This is expected for itinerant electron motion along a chain with 
parallel spins. 
 However,  we also see the development of new populations
 with $M= 1$ and $m=1/2$, primarily at the corner sites, Fig. \ref{fig1}(h) and (j).
 These spin populations indicate local spin canting away from the $z$--axis and the pristine AFM magnetic order. 
Such quantum spin canting is seen  for  the full range of magnetic field and lattice displacements
in Fig. S1.
For  $t > T$, all populations have reached stationary values within numerical accuracy (Fig. S2).

\begin{figure}[t]
\begin{center}
\includegraphics [scale=0.45] {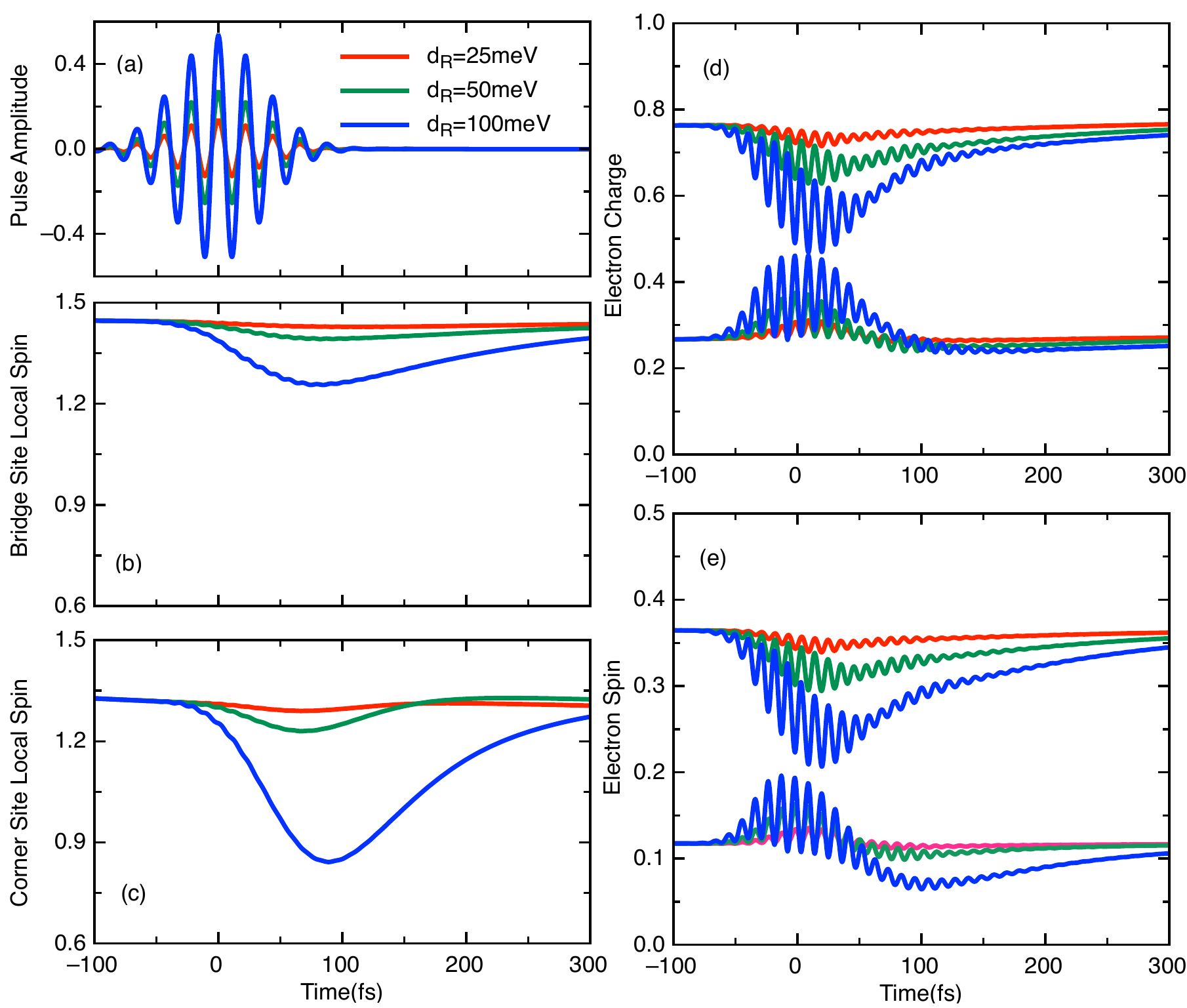}  
\caption{(Color online)  Non-adiabatic dynamics driven  by  the oscillating electric fields shown  in (a).
(b) and (c): Bridge and corner site local spin dynamics $S_z(t)$.  
(d) and (e): Itinerant electron charge and spin dynamics
at bridge and corner sites. 
}\label{fig_chg+spin-dynamics}
\end{center}
\end{figure}

\textit{Non--adiabatic dynamics}.---
We now 
consider the  non--adiabatic dynamics driven by time--periodic modulation 
of the electronic hopping amplitudes during electric field oscillations, Fig. 1(b).    
Fig. \ref{fig_chg+spin-dynamics}(a)
shows the multi--cycle electric fields considered here, 
with duration $t_p= 100$fs and frequency $\omega_p$ close to the inter--site energy barrier. 
These laser fields  drive 
(i) inter--site coherences with 
dephasing times $T_2 \sim$ 20fs ($T_2 < t_p$) 
that characterize  the inter--site electronic coherent hopping,    
and  (ii) non--thermal charge and spin coherent 
 populations with  lifetime  $T_1 \sim 200$fs
comparable to $t_p$.
   The predicted effects are enhanced for 
   longer $T_1$ and $T_2$.
The   laser electric field
introduces
a transient modulation of the 
 hopping amplitude
$V_{\alpha \beta}(i-j)$  
between neighboring sites during few cycles of lightwave oscillations, 
described, e.g., 
by using the Peierls substitution  (Supplementary Section S4). 
Thus the lightwave  
accelerates itinerant  electrons across the lattice.
We characterize  the electromagnetic  coupling strength by the ``Rabi energy"  
 $d_R$=$e a E$,
where $a$ is the lattice constant, $e$ the electron charge, and $E$ 
the electric field amplitude.
The initial condition for laser--driven dynamics  is the 
stationary  state after adiabatic turn-on of electronic hopping, 
Fig. \ref{fig1}.  
The light--induced quantum transport
during cycles of electric field oscillations  
  is described by the time evolution of the 
off--diagonal density matrix elements between all 
 pairs of lattice sites $(i,j)$. 
Fig. \ref{fig_chg+spin-dynamics}  shows the resulting  local spin 
driven dynamics, $S_{z}(t)$,
at the bridge, Fig. \ref{fig_chg+spin-dynamics}(b), and corner, Fig \ref{fig_chg+spin-dynamics}(c), lattice sites.  
The difference in   
local spin changes  between the two  sites results  in 
 femtosecond magnetization for 
a small  magnetic field $B_{ext}$ that breaks the symmetry.  
This magnetization is determined by the spin--resolved 
population  dynamics,
shown in Fig. \ref{fig_chg+spin-dynamics}(d).
The  charge imbalance between 
bridge and corner sites in the inital state gives way to a uniform charge distribution,
 which   relaxes back to equilibrium after $T_1$ if we 
assume frozen lattice displacements (more on this later). 
$S_z(t)$  in 
Figs \ref{fig_chg+spin-dynamics}(b) and (c)
 decreases from equilibrium 
at both bridge and corner sites, which signifies
quantum spin canting  
with respect to the initial AFM orientation along the $z$--axis
during ightwave   quantum transport.
Fig.  \ref{fig_chg+spin-dynamics}(e) shows  the photoinduced itinerant electron spin, which 
 drives the above local spin  canting
 via the off--diagonal onsite magnetic interaction 
$\propto {\bf S}^{-} \cdot {\bf s}^{+}$. 
The  time-evolution 
 displays  oscillations  with frequency $2\omega_p$
(Supplementary Figure S3) that  reflect the 
coherent nature of the 
inter--site spin and charge transfer  
during lighwave cycles.

\begin{figure}[t]
\begin{center}
\includegraphics [scale=0.46] {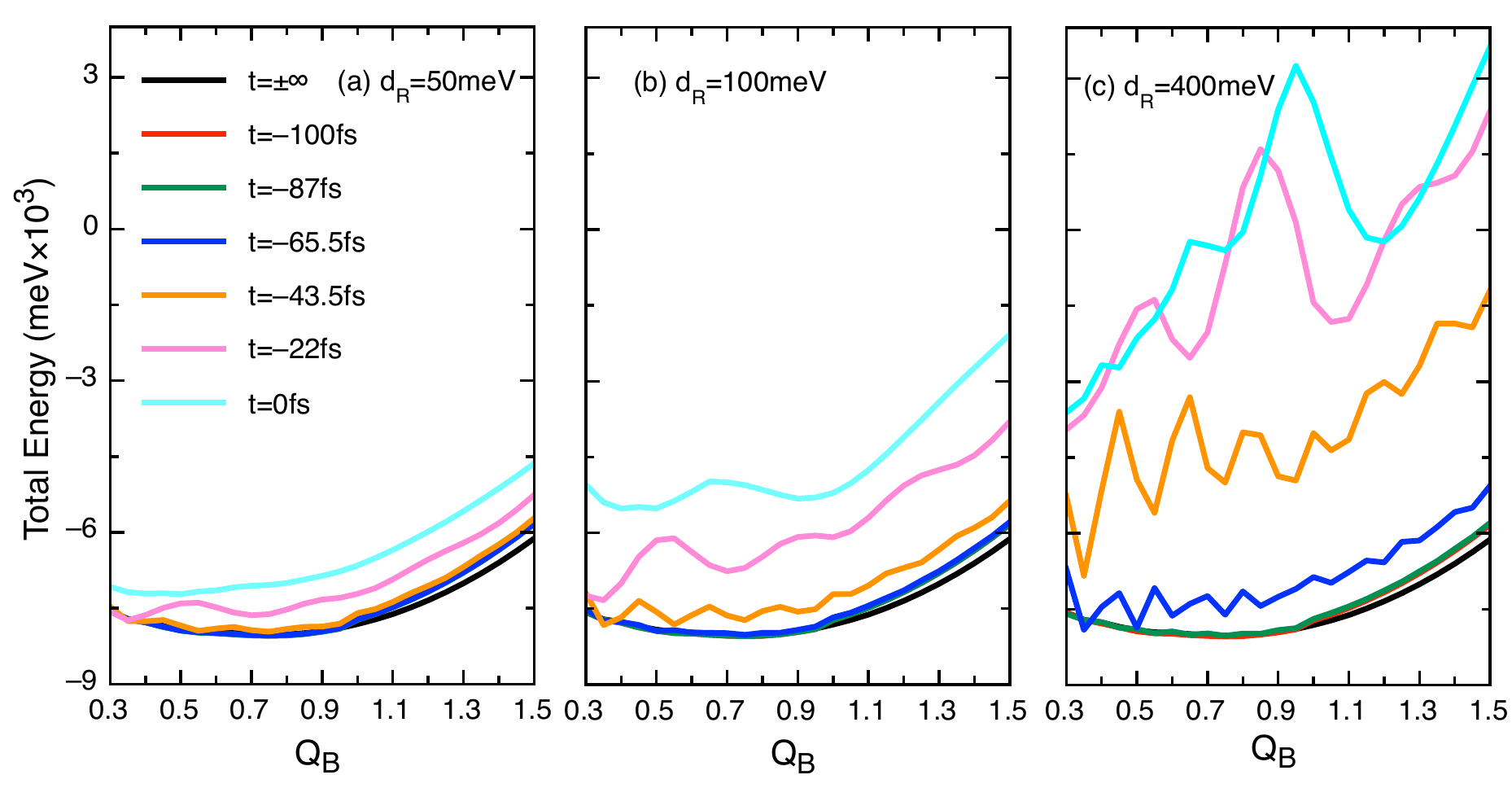}  
\caption{(Color online) Lightwave driven  insulator to metal transition: 
Total energy landscape at characteristic time instances during cycles of lightwave oscillations 
for low (a), intermediate (b), and high (c) Rabi (Zener tunneling) energies $d_R$=$e E a$.
}\label{fig_EnergyLandScape}
\end{center}
\end{figure}

%

To elucidate  the light--driven 
quantum spin fluctuations,
we compare  in Fig. \ref{fig1}(e,g,h,k)
 the spin--resolved  populations of the different configurations 
 $| i m \rangle$
and $| i \alpha M\rangle$  
 at bridge (panels (e) and (g)) and corner (panels (i) and (k)) 
 lattice sites.
At the bridge sites,
 the $M=2$ majority population 
 decreases from its  ground state value, Fig. \ref{fig1}(g), 
 with a simultaneous 
increase in the $m=3/2$ local spin population, Fig. \ref{fig1}(e). 
This reflects the coherent  hopping 
of 
an itinerant electron FM--coupled  to 
the local spin 
  from a
 bridge to a corner site
 during electric field cycles of oscillation.  
At the same time, 
 new time--delayed local spin populations
with $m<3/2$ 
develop at the bridge sites, Fig \ref{fig1}(e),
which signifies non--instantaneous 
quantum  canting of the local spin 
away from its  equilibrium  orientation 
$m$=3/2. Moreover, 
the population of $M<$2
 states, Fig. \ref{fig1}(g),  comes from  
electronic hopping back 
to the bridge sites.
Quantum spin canting  
is stronger on the corner sites. This is 
seen in
Figs \ref{fig_chg+spin-dynamics}(i--k)
by the significant population of $m <$3/2 and $M <$2 configurations. 
This difference between corner and bridge sites leads to the development of  
 FM correlation 
during lightwave cycles of oscillation, which arises from 
coherent hopping between sites with AFM spins
simultaneously with   quantum spin canting. Such lightwave quantum transport of spin and charge 
populates different spin states 
prior to relaxation ($T_1$) while  
simultaneously resulting in a more uniform  charge distribution, $\Delta n \sim 0.5$. 

So far, we have assumed that the  lattice displacements are frozen 
during electronic hopping  timescales. 
Fig. S1(d)  shows that this approximation is justified for adiabatic turn--on of $V(t)$. 
as 
the total energy 
landscape $E(Q_B)$  
is not influenced significantly by the electronic charge transfer.
In contrast,  Fig. \ref{fig_EnergyLandScape} shows that lightwave 
quantum transport of electron spin and charge significantly changes the energy 
landscape  out of  equilibrium. Fig. \ref{fig_EnergyLandScape}(a) shows  the changes in 
total energy $E(Q_B,t)$ due to the light--driven spin--charge populations and coherences . Unlike for adiabatic 
hopping, 
the electronic quantum  transport  tuned by $d_R$ 
leads to significant 
changes in the overall shape of E($Q_B,t)$ during cycles of oscillation
(Fig. 3). 
Below electric field threshold, 
$d_R=50$meV in Fig.  3(a), 
the photoexcited system is  insulating,
since the
total energy minimum remains at a finite $Q_B <Q^{eq}_B$.
The main effect with increasing $d_R$  is  the softening of the  phonon mode,  as well as a non--parabolic dependence of E($Q_B)$, whch are evident    for 
 $d_R=100$meV after three cycles of oscillations  
(Fig. \ref{fig_EnergyLandScape}(b)).  
Such effects of 
lightwave charge and spin coherent populations
induce  anharmonic 
lattice nonlinear motion and forces. 
Above threshold,
$d_R$=400meV, 
Fig. \ref{fig_EnergyLandScape}(c) shows that a new global minimum at $Q_B$=0  develops
 during lightwave cycles. 
 This  change in the 
non--equilibrium total energy  shape
favors a  metallic  phase 
not possible in equilibrium.  
The  dynamics of the phase transition,  driven  by  coherent lattice displacements  $Q_B(t)$ due to  time--dependent forces  $-\frac{dE(Q_B,t)}{d Q_B} $ \cite{lingos-17},
will be considered elsewhere.

In conclusion, 
  inter--site excitations of itinerant electron spin and charge 
interacting strongly with 
   an AFM spin   background during cycles of 
  oscillations of the modulated 
  inter--site coherent hopping amplitude
(i) Create a more  homogeneous metallic--like nonthermal electronic population
 throughout the lattice, 
(ii) Drive transient magnetization from AFM--ordered spins 
 via quantum spin fluctuations,
and 
(iii) Destabilize the AFM/insulating phase 
with lattice displacements $Q_i \ne 0$ 
towards a transient metallic phase 
with $Q_i  \sim 0$. 
Above electric field threshold, 
these non--equilibrium effects 
create  non--thermally an  initial condition for 
nonlinear lattice dynamics, Fig. \ref{fig_EnergyLandScape}, 
by drastically modifying the  energy  landscape in ways not possible 
close to equilibrium.
The latter change in energy landscape makes the effects 
calculated for frozen lattice much more pronounced in a self--consistent calculation.
We can thus envision ultrafast manipulation of 
an insulating  phase with non--collinear spins by tunable 
laser pulse sequences, which 
can remove lattice distortions 
and coherently drive  insulator--to--metal phase transitions 
simultaneously with transient magnetization  
via  strong spin--charge quantum couplings.

This work was supported  
by the US Department of Energy under contract \# DE-SC0019137
and was made possible in part by a grant for high performance computing resources and technical support
 from the Alabama Supercomputer Authority (ASA).
J.W. was supported by the Ames Laboratory, the US Department of Energy, Office of Science,
Basic Energy Sciences, Materials Science and Engineering Division under contract No. DEAC02-
07CH11358 (data analysis).

\end{document}